\documentstyle[12pt]{article}
\begin{document}

\begin{flushright}
RU--97--92 \\
\today 
\end{flushright}

\begin{center}
\bigskip\bigskip
{\Large \bf  On the Effective Action of  $N=1$ Supersymmetric
Yang-Mills Theory}
\vspace{0.3in}      

{\bf G.R. Farrar,~G. Gabadadze,~M. Schwetz}
\vspace{0.2in}

{\baselineskip=14pt
Department of Physics and Astronomy, Rutgers University \\
Piscataway, New Jersey 08855, USA}\\
{\rm emails: farrar, gabad, myckola@physics.rutgers.edu}

\vspace{0.2in}
\end{center}

\vspace{0.9cm}
\begin{center}
{\bf Abstract}
\end{center} 
\vspace{0.3in}
We propose a generalization of the Veneziano-Yankielowicz effective
low-energy action for $N=1$ SUSY Yang-Mills theory which includes
composite operators interpolating pure gluonic bound states.
The chiral supermultiplet of anomalies is embedded in a larger three-form
multiplet and an extra term in the effective action is introduced.  The mass
spectrum  and mixing of  the lowest-spin bound states are studied within
the effective Lagrangian approach.  The physical mass eigenstates form two
multiplets, each containing a scalar, pseudoscalar and Weyl fermion.  The
multiplet containing the states which are most closely related to glueballs
is the lighter one. 
\vspace{2cm}


\newpage
\noindent {\bf Introduction}

Nonabelian supersymmetric gauge theories have attracted  much attention
since tremendous progress was made in understanding
the ground state structure of some of those models \cite {S},
\cite {SW}. And yet some puzzles remain.  In the present paper we consider
the simplest nonabelian supersymmetric gauge model, $N=1$ SUSY Yang-Mills
theory (SYM) with the $SU(N_c)$ gauge group. This model describes interactions
of gluons and gluinos. In analogy with QCD, one expects that the spectrum
of the model consists of colorless bound states of those fundamental
excitations, namely glueballs ($gg$), gluinoball-mesons ($\tilde{g} \tilde{g}$),
and "glueballinos" ($g \tilde{g}$). 

Description of the color singlet bound states in terms of nonabelian gauge
fields is a complicated task. However, one can use  the effective field
theory technique. Knowing all global symmetries and anomalies of the model one constructs
an effective action in terms of colorless variables.  For the case of
$N=1$ SUSY Yang-Mills theory, the effective action\footnote[1]{To what extent
that action can be thought of as describing particles and can be called the
effective action  will be discussed  in the next section.} was constructed
by Veneziano and Yankielowicz (VY) \cite {VY}. The VY action \cite {VY}
involves interpolating operators for gluino-gluino and gluino-gluon bound
states.  However, it does not include composite operators corresponding to
pure gluonic composites (glueballs) and attempts to generalize the VY action
to include them have failed up to now.  This is perplexing since QCD is
closely related to SYM.  If there were some fundamental impediment to
constructing a supersymmetric effective action representing the full
set of expected colorless fields (glueballs as well as gluinoball
mesons), it could have important implications for our understanding of the
QCD spectrum.

In this paper we report that it is possible to extend the VY action in a
way that allows interpolating operators corresponding to the  gluon-gluon
bound states to be included as dynamical variables. 
In section 1 we briefly discuss the VY approach and indicate 
why the interpolating operators of the gluon-gluon bound states 
are not included in that action. We show, in section 2,  that the  problem
can be cured by embedding the chiral multiplet, which is used in the VY
construction, into a larger  tensor multiplet.  Rewriting the VY action in
terms of that tensor multiplet, and adding one extra  term to the action,
one discovers that the low energy theory includes the interpolating
operators of all the  lowest-spin bound states which are expected on the
basis of the naive ``valence'' construction including $l=0$ and $l=1$ states.
The spectrum is of course consistent with the $N=1$ SUSY algebra.
In section 3 we study the mass spectrum and mixing of gluino-gluino,
gluon-gluino and gluon-gluon composites.  We briefly compare our results
with recent lattice gauge theory calculations of the spectrum 
of $N=1$ SUSY Yang-Mills  theory.
\vspace{0.3in}\\
{\bf 1. The VY effective action }

The classical action of  $N=1$ SYM theory is invariant  
under $U(1)_R$, scale 
and superconformal transformations. 
In the 
quantum theory these symmetries are 
broken by the  chiral, scale and superconformal  anomalies respectively.
Composite operators that appear  in the expressions for the 
anomalies can be thought of as  component fields  
of a chiral supermultiplet $S$ \cite {WessZumino}
\begin{eqnarray}
S\equiv A(y)+\sqrt{2} \theta \Psi(y) + \theta^2 F(y), \nonumber
\end{eqnarray}
where the following notations for the composite operators
are  introduced\footnote[1]{We follow
conventions of Wess and Bagger \cite {WessBagger}.}
\begin{eqnarray}
A\equiv { \beta (g) \over 2 g} \lambda^{\alpha}\lambda_{\alpha},~~~~~
\sqrt{2}\Psi_{\alpha} \equiv - { \beta (g) \over 2 g} 
\{ -i \lambda_{\alpha} D+(\sigma^{\mu\nu} \lambda)_{\alpha}G_{\mu\nu} \},
\nonumber  \\
F\equiv  - { \beta(g) \over g} \{ -{1\over 4}G^2_{\mu\nu}-
{i\over 2} {\bar \lambda} {\bar \sigma} 
{\overline  \nabla  }\lambda+{1\over 2}D^2-
{i\over 4} G_{\mu\nu}{\tilde G_{\mu\nu}}+{i\over 2} \partial_\mu
J_\mu^5 \}.  
\label{Composites}
\end{eqnarray}
In these expressions 
$\beta(g)$ stands for the SYM beta function for which the 
exact expression is  known  \cite {beta}, 
$\lambda_\alpha$
denotes a two-component gluino field (Weyl spinor), $D$ 
stands for the $D$ component of the nonabelian vector superfield,
and the normalization of the dual stress-tensor is given by
${\tilde G_{\mu\nu}}={1\over 2} \varepsilon_{\mu\nu\lambda\tau} 
G^{\lambda\tau}$.
All the composite operators in eq. (\ref {Composites}) 
have zero anomalous dimensions. 
Thus, they can be  treated as interpolating
operators for  the lowest-spin  colorless bound states present in the 
spectrum of the model. Furthermore, one might argue that 
the effective action for those
bound states  can  be specified  in terms of the $S$ field
\cite {VY}. 

Following ref. \cite {VY} one constructs  
an effective superpotential for the theory.  
The superpotential  
which reproduces   correctly  all the three anomalies of SYM 
theory  
is given by  the expression  \cite {VY}
\begin{eqnarray}
W(S) \propto ( S \log {S\over \mu^3}-S)+ {\rm h.c.}. 
\label{Superpotential}
\end{eqnarray}
Here $\mu$ is the dynamically generated  scale of SYM theory:
$\mu = \mu_0 e^{-{8\pi^2 \over 3N_c g^2}}$ where the running coupling
$g$ is defined at some scale $\mu_0$. 

In order to fix a complete  effective Lagrangian description  
of the lowest-spin  states of the theory one needs 
to define  also an effective  K\"ahler potential  
in terms of the $S$ field. 
Since all the anomalies are already taken into account by 
the superpotential  (\ref {Superpotential}), 
the K\"ahler potential should in its turn respect the $U(1)_R$, 
scale and  superconformal symmetries. As a simplest expression satisfying
those requirements one finds \cite {VY} 
\begin{eqnarray}
K_{\rm VY}(S^+,S)\equiv (S^+S)^{1/3}. 
\label{K}
\end{eqnarray}
The most general form of the 
K\"ahler potential as a function of $S$, $S^+$, and 
derivatives of these superfields 
was determined in ref. \cite {Shore}. 
It was shown \cite {Shore} that 
the only expression that satisfies chiral, scale and superconformal
Ward identities can be written in the following form 
$$ 
K(S^+,S)\equiv (S^+S)^{1/3}f(\chi, \chi^+),~~~{\rm where }~~~
\chi\equiv S ^{1/3} ({\bar D}^2 S^{+1/3})^{-1/2}.
$$ 
Here $f$ stands for an arbitrary
function of two variables satisfying the reality condition 
$f^*(x,y)=f(y,x)$ \cite {Shore}. 
However, the  K\"ahler potential $K$,  being combined with the superpotential
(\ref {Superpotential}),   
yields   a  meaningful theory with a bounded-from-below potential if
and only if $f=1$ \cite {Shore}. 
Thus, the VY  anzatz  (\ref {K}),  being the
simplest one, turns out to be  the only physically acceptable  expression
which could  be combined with the superpotential (\ref {Superpotential}) 
to define an effective action of  the model. 
Let us stress again that the validity of assertion assumes 
that the effective action of the theory is defined 
as a functional of  the $S$ (and $S^+$)  field  
and its higher derivatives only.

Bringing eqs. (\ref {Superpotential}) and (\ref {K}) together,
one writes down the VY effective action in the following form 
\begin{eqnarray}
{\rm Action}=\int  d^4x~~{1\over \alpha} (S^+S)^{1/3}|_D+ 
\gamma [( S \log {S\over \mu^3}-S)|_F+{\rm h.c.}],
\label{Action}
\end{eqnarray}
where the positive constants  $\alpha$ and $\gamma$ are  introduced.
The value of  $\gamma$  can be fixed explicitly \cite {Seiberg}. In
our notation $\gamma = -(N_c g/16 \pi^2 \beta (g))>0$. 

The axial $U(1)_R$ symmetry is broken by the  anomaly to 
a discrete ${\bf Z}_{2N_c}$ symmetry in the $N=1$ $SU(N)$ SYM theory. 
That ${\bf Z}_{2N_c}$ symmetry group is itself broken down to 
${\bf Z}_{2}$  due to the nonzero gluino condensate 
\cite {Condensate}:
\begin{equation}
\label{cond}
\langle \lambda \lambda \rangle \propto  \mu^3 ~
e^{{2\pi i k \over N_c}}~,~~~~~~~ k = 0,1,...,N_c-1~.
\end{equation}
Hence we have $N_c$ physically inequivalent vacua , each characterized by
its own phase of the gaugino condensate (\ref{cond}).
The VY effective action of (\ref{Action}) describes the theory around
one of them (here $k=0$). 
It was conjectured   in ref. \cite{KShifman}  
that the theory might contain a new vacuum
which is in a ${\bf Z}_{2N_c}$ chirally symmetric superconformal phase
with zero gaugino condensate. A number of arguments have been given against 
the existence of such a phase \cite{mm}, \cite{cm}.
In any case, we concentrate our attention on  the conventional 
phase of the theory with nonzero gluino condensate \cite {Condensate}.

At this stage  we would like to comment on the  physical meaning of the 
effective action (\ref {Action}). This is not an  effective action 
in the Wilsonian sense (see discussions  in refs. \cite {Shore}, \cite {SVZ}). 
In ref. \cite {Shore} the action (\ref {Action})
was constructed  as a generating functional of one-particle-irreducible (1PI)
Green's functions, an object  first introduced in Quantum Field Theory
in ref. \cite {GoldstoneSalamWeinberg}. 
That means that  the action (\ref {Action}), being written in
terms of composite colorless fields  of  SYM theory,  can  be used 
to calculate  various Green's functions of those composite variables
\footnote{One should keep in mind that actual variables
in this action are VEV's of the composite operators calculated
in a theory with nonzero external sources for those operators
(see for instance discussions  in ref. \cite {Shore}). 
For the sake of simplicity of presentation we denote those 
VEV's by the corresponding composite fields.}. 
Performing those calculations, however, one is not supposed to 
take into account
diagrams  with  composite fields  being propagating in  virtual loops.
Loop effects  are  already included 
in effective  vertices and
propagators  occuring in the action  (\ref {Action}). 
Thus, all calculations 
with the expression  (\ref {Action})  are to be carried out 
in  the  tree level approximation 
\footnote[3]{For non-supersymmetric gluodynamics, analogous actions
were constructed in ref. \cite {Schechter}, \cite {MigdalShifman} for
the CP even sector of the theory,  and in ref. 
\cite{Gabad} for the CP odd sector of the model.}. 

The simplest kind   of  Green's function  one might be 
interested in is a two point correlator. As we mentioned above, 
the composite operators
entering the expression (\ref {Action}) are the interpolating 
fields for the bound states of $N=1$ SYM theory. 
Thus, a two point correlator (or simply a 
propagator)  of those fields can be used to determine  the
mass of the corresponding bound state. 
Hence, the effective action 
(\ref {Action}) (or more exactly the generating functional of 1PI diagrams)
can readily be used to deduce masses of composite bound states of the 
theory. In what follows we concentrate our attention on 
this aspect of the effective action approach to $N=1$ SUSY YM theory. 

A great advantage of SUSY gauge theories is 
the fact that all physical states of the model are    
components of SUSY multiplets. That can be deduced directly from the
corresponding SUSY algebra.  Based on the $N=1$ SUSY algebra written in terms
of generators with  a definite space-time parity \cite {OgievetskyMezincescu}
one expects the following
lowest-spin multiplets of spin-parity eigenstates:
$[0^{-+},~~{1\over 2}^{i+},~~0^{++}]$ and $[0^{++},~~
{1\over 2}^{(-i)+},~~0^{-+}]$. 
In SYM theory one expects these states to
be realized as the following composites:

${\rm Ia}$~  A pseudoscalar, $0^{-+}$,  $l=0,~s=0$ 
gluino-gluino bound state; 

${\rm Ib}$~   A spinor,  ${1\over 2}^{i+}$,  $l=1,~s=1/2$ gluon-gluino bound 
state; 

${\rm Ic}$~  A scalar,  $0^{++}$,   $l=1,~s=1$  gluino-gluino bound state.

${\rm IIa}$  A scalar,  $0^{++}$,  $l=0,~s=0$  gluon-gluon bound state;

${\rm IIb}$  A  spinor, ${1\over 2}^{(-i)+}$ $l=0,~s=1/2$ gluon-gluino bound 
state;

${\rm IIc}$ A  pseudoscalar, $0^{-+}$ $l=1,~s=1$ gluon-gluon  bound state 
\footnote[6]{Though  the spin-orbital  decomposition
is  an intrinsically nonrelativistic notion, it can  be applicable
to relativistic cases (see the discussion in ref. \cite
{Landafshitz}). A ``spin'' in that case is defined as the rank of the
spinor which  describes the corresponding wavefunction, and  an ``orbital 
momentum'' is set by the coordinate dependence of the relativistic
wave function \cite {Landafshitz}.}.  

In general, these states would be assigned to  two different
supermultiplets. 
Note that  the complex 
fields $A$, $\Psi$ and $F$, introduced in (\ref {Components}),  form linear
combinations of the interpolating operators for the states listed above.
So at least formally,  the above bound states are present 
in  the supermultiplet $S$ and, consequently, in the 
action (\ref {Action}).
Hovewer, 
the gluon-gluon bound states, referred to hereafter as ``glueballs'', 
enter the action (\ref {Action})  through  the $F$ term of 
the chiral multiplet. $F$ terms generically appear as auxiliary
fields of a  model and are usually  integrated out. 
For instance in the VY approach 
there is  no kinetic term  for the $F$ component  
of the chiral superfield so  the $F$ 
field can be integrated out by means of  
equations of motion.  Having eliminated the $F$  field
one is left with the effective Lagrangian description 
of the gluino-containing bound states only (the first three  states in the
list above).  Thus no  glueballs are present 
in the VY action\footnote[7]{When the equations of motion are used 
the number of degrees of freedom of the fermionic
field $\Psi$ also reduces. The off-mass-shell spinor $\Psi$ has 
four real degrees of freedom, while it propagates only two independent
degrees of freedom when  the on-shell condition is imposed. 
Those two degrees of freedom describe only one (out of two) 
gluino-gluon bound states given  in the list above.}. 

In general,  there is no reason to believe 
that in SYM theory the
spin-zero glueball states are 
heavier than the gluino containing mesons. 
Thus, at least $ a~priori$,  spin-zero glueballs have every right to be 
considered  in the effective action of the model. 

One can attempt to construct a new chiral superfield which
would include the  interpolating operators for 
glueballs  in a  lowest supercomponent \cite {NSM}.
An appropriate superfield is  proportional to 
${D^2 S}$ \cite {NSM}. However,  R-symmetry arguments do
not allow one to introduce a nontrivial coupling 
of that chiral multiplet to the VY supermultiplet \cite {NSM}. 
Another approach is needed. 
\vspace{0.3in}\\
{\bf 2. Generalization of the VY effective action }

In order to determine  how glueballs can be included in the action 
(\ref {Action}) 
let us concentrate our attention on the expression for the $F$ field.
Using the equation  of motion for the gluino  field and for the $D$
component one gets\footnote[5]{In general, one is not allowed to use 
the equation of motion if the VEV of the $F$ field is cosidered.} 
\begin{eqnarray}
F\equiv { \beta (g) \over4  g}[G^2_{\mu\nu}
+i G_{\mu\nu}{\tilde G_{\mu\nu}}]. \nonumber 
\end{eqnarray}
Let us introduce the following notations 
\begin{eqnarray}
\Sigma \equiv { \beta (g) \over4  g} G^2_{\mu\nu}, ~~~~~~~
Q\equiv { \beta (g) \over4  g} G_{\mu\nu}{\tilde G_{\mu\nu}}.\nonumber
\end{eqnarray}
Thus, the decomposition of the $F$ field into  its real and imaginary
parts  is 
$ F=\Sigma+i Q$. 
As we have already mentioned,  
the $F$ field appears in the VY action without
a kinetic term. The term  bilinear in  the $F$ field 
is proportional to 
 \begin{eqnarray}
F^+F=\Sigma^2+Q^2. \nonumber
\end{eqnarray}
Besides that,  there are terms linear in  the $F$ field 
in the expression for the effective action, thus, 
the $F$ field can easily be integrated out 
by means of its  algebraic equations of motion \cite {VY}. 

In order to reveal 
subtleties of this procedure let us write down the 
following relation
\begin{eqnarray}
Q={1\over 4! } \varepsilon_{\mu\nu\alpha\beta} H^{\mu\nu\alpha\beta}, 
\label{QC}
\end{eqnarray}
where $ H^{\mu\nu\alpha\beta}$ is a field strength 
for a three-form potential $C_{\nu\alpha\beta}$, 
$
H_{\mu\nu\alpha\beta}=\partial_\mu C_{\nu\alpha\beta}-
\partial_\nu C_{\mu\alpha\beta}-\partial_\alpha C_{\nu\mu\beta}-
\partial_\beta C_{\nu\alpha\mu}.  \nonumber 
$
The $C_{\mu\nu\alpha}$ field  itself is defined as a   
composite operator of colored gluon fields $A^a_\mu$, 
$
C_{\mu\nu\alpha}={\beta (g) \over 64 g \pi^2}(A^a_\mu 
{\overline {\partial}}_\nu
A^a_\alpha-A^a_\nu {\overline {\partial}}_\mu A^a_\alpha-A^a_\alpha 
{\overline {\partial}}_\nu
A^a_\mu+ 2 f_{abc}A^a_\mu A^b_\nu  A^c_\alpha), \nonumber
$
with $f_{abc}$ being  structure constants of the corresponding $SU(N_c)$
gauge group.  The right-left  derivative in this expression acts  as 
$A{\overline {\partial}}B\equiv A (\partial B)-(\partial A) B $
\footnote[8]{The quantity $Q$ 
can also be expressed through the Chern-Simons
current $K_\mu $ as   $Q=\partial_\mu K_\mu $. Using 
this equation  one can deduce  the relation between the Chern-Simons
current and the three-form potential $C_{\nu\alpha\beta}$, these two
quantities are  Hodge dual to each other:
$K^{\mu}={1\over 3!}\varepsilon^{\mu\nu\alpha\beta}C_{\nu\alpha\beta}$.}.
 
Using these definitions 
one finds that the expression bilinear in  the $F$ field  acquires  the 
following form 
\begin{eqnarray}
F^+F=\Sigma^2-{1\over 4! }  H^2_{\mu\nu\alpha\beta}. \nonumber 
\end{eqnarray}
The second term in this expression is a 
kinetic term  for the three-form potential $C_{\mu\nu\alpha}$.
As before,  the $\Sigma$ field can be integrated out, 
however one should   be
careful in dealing with the $C_{\mu\nu\alpha}$ field. 

In ref. \cite {Gabad} it was  argued that  the 
three-form field $C_{\mu\nu\alpha}$ plays an important role 
in the description of the  pseudoscalar glueball. 
That glueball can be coupled to the QCD $\eta'$ meson by means of the 
$C_{\mu\nu\alpha}$ field \cite {V}.  In the case of SYM theory 
the analog of the $\eta'$ meson is the gluino-gluino 
bound state which acquires mass due to the anomaly in the 
$U(1)_R$ current within the VY approach. 
Thus,  it is natural to  attempt to 
couple the pseudoscalar glueball to the pseudoscalar gluino-gluino 
bound state within the VY action 
using the three-form potential $C_{\mu\nu\alpha}$.

To elaborate this approach,   let us rewrite the SUSY transformations 
for the components of the $S$ superfield in terms of $\Sigma $
and $C_{\mu\nu\alpha}$ (instead of $F$ and $F^+$) \cite {Gates}, 
\cite {OvrutWaldram}
\begin{eqnarray}
\delta_{\zeta} A= \sqrt{2} \zeta \Psi,~~~~~~~~~
\delta_{\zeta}\Psi =i\sqrt{2}\sigma^{\mu}{\bar \zeta}\partial_\mu A
+\sqrt{2}\zeta (\Sigma +{i\over 6}\varepsilon_{\mu\nu\alpha\beta}
\partial^\mu C^{\nu\alpha\beta}), \nonumber \\
\delta_{\zeta}\Sigma = {i\over \sqrt{2}}({\bar \zeta}{\bar \sigma}^\mu
\partial_\mu\Psi+ \zeta \sigma ^\mu \partial_\mu {\bar \Psi}),
~~~~~
\delta_{\zeta}C_{\nu\alpha\beta}={1 \over \sqrt{2}}
\varepsilon_{\nu\alpha\beta\mu }({\bar \zeta}{\bar \sigma}^\mu
\Psi- \zeta \sigma ^\mu  {\bar \Psi}). \nonumber
\end{eqnarray} 
The set of fields given  above forms  an irreducible
representation of supersymmetry algebra. All these
fields can be assigned to a supermultiplet 
introduced in ref. \cite {Gates}. 
That supermultiplet is called a constrained three-form supermultiplet 
\cite {Gates},  \cite {GatesGRZ}. 
The easiest way to present this multiplet
is to introduce the following real tensor 
superfield $U$ \footnote{In this
discussion we follow the conventions of ref. \cite {OvrutWaldram}.}
\begin{eqnarray}
U=B+i\theta \chi -i {\bar \theta} {\bar \chi}+{1\over 16}\theta^2 {A^*}+
{1\over 16} {\bar \theta}^2 A+{1\over 48 }\theta \sigma^\mu {\bar
\theta} \varepsilon_{\mu\nu\alpha\beta}C^{\nu\alpha\beta}+ 
\nonumber \\
{1\over 2} \theta^2 {\bar \theta} \left ( {\sqrt{2} \over 8}{\bar
\Psi} +{\bar \sigma}^\mu \partial_\mu \chi \right )+
{1\over 2}{\bar  \theta}^2 \theta  \left ( {\sqrt{2} \over 8}
\Psi - \sigma ^\mu \partial_\mu {\bar \chi }\right )+{1\over 4}
\theta^2 {\bar \theta^2} \left ( {1\over 4} \Sigma -\partial^2 B\right ).
\label{U}
\end{eqnarray} 
It is a matter of a straightforward calculation to check that the 
real superfield $U$  satisfies the relation\footnote{Despite a 
seeming similarity, the tensor multiplet $U$ should not be
interpreted as a usual vector multiplet. The vector field which might
be introduced in this approach as a Hodge dual of  the three-form 
potential $C_{\mu\nu\alpha}$ would give
mass terms with the wrong sign in our approach (see section 2),
thus,  the actual physical variable  is the three-form potential  
$C_{\mu\nu\alpha}$ rather than its dual vector field 
(the Chern-Simons current).} 
\begin{eqnarray}
{\bar D}^2 U=-{1\over 4} S. 
\label{US}
\end{eqnarray} 
Thus, the real tensor multiplet $U$, defined by the 
expression (\ref {US}),  includes  all the 
components of the chiral supermultiplet $S$. Besides that 
the multiplet has also an additional
scalar $B$ and  fermion $\chi$. Thus, using the relation (\ref {US}) 
the VY action can be rewritten
in terms of the bigger multiplet $U$. We will show below that this 
allows one to include glueball operators in the effective
action\footnote{This representation of SUSY was earlier used in
the context of SYM theory in ref. \cite {Derendinger} to study
the phenomenon of gaugino condensation with a field dependent 
gauge coupling. We thank J.-P. Derendinger for bringing ref.
\cite {Derendinger} to our attention.}. 

First, let us  notice some  features of 
SUSY transformations of the components of the $U$ field. The
components which are shared by the tensor multiplet $U$  
and the chiral
multiplet $S$ (namely $A,~\Psi,~\Sigma~~ {\rm and}~~C$ ) transform 
among themselves, while  other fields ($B$
and $\chi$) are connected by SUSY rotations to the other four components. 
Furthermore, one can define  a  ``gauge'' transformation of 
the $U$ field as the following shift $U\rightarrow U+Y$, where the superfield
$Y$ satisfies the relation ${\bar D}^2 Y=0$. 
It is important to notice that by means of 
this `` gauge '' transformation one can get rid of the 
$B$ and $\chi$ fields in the expression for the $U$ multiplet.
This is the analog of the Wess-Zumino gauge for the tensor multiplet $U$. 
Thus, any Lagrangian
written  in terms of the $S$ field only, if reexpressed
 in terms of the $U$ field,
is necessarily invariant under the ``gauge'' 
transformation defined above. 
As a result, the $B$ and $\chi$ fields can  always be ``gauged''
away from that Lagrangian. Thus in order to be able to retain 
the $B$ and $\chi$   fields  as dynamical variables  
one must include terms in the Lagrangian which 
breaks this ``gauge'' invariance. The simplest term of this type is the
quadratic term $(U^2)|_D$. Once such a term is included in the
Lagrangian, the ``gauge'' symmetry becomes explicitly broken and the 
$B$ and $\chi$  components of the superfield $U$ survive as 
dynamical variables. 

Let us now apply the $U$ field formalism to the VY action.
In the case at hand  the chiral symmetry is spontaneously broken by
the gluino condensate. In terms of the chiral superfield this
corresponds to the existence of a nonzero VEV of the $S$ field
\begin{eqnarray}
\langle S\rangle =\mu^3. \nonumber 
\end{eqnarray}
With that in mind the  appropriate relation between the $U$ field and 
the chiral multiplet is
\begin{eqnarray}
{\bar D}^2 U=-{1\over 4} (S-\langle S\rangle ). 
\label{USrelation}
\end{eqnarray}
The only result of this modification is that the field $A$
in eq. (\ref {U}) gets replaced by the quantity 
$A-\langle A \rangle$.

Now use the relation (\ref {USrelation})
to rewrite the action (\ref {Action}) 
in terms of the $U$ field.
In order to break the ``gauge'' invariance of the VY action we
add  a  term proportional to $U^2$ to the VY Lagrangian. 
An  appropriate  term with zero  R-charge and correct  
dimensionality  is
\begin{eqnarray}
\left ( -{U^2\over (S^{+}S)^{1/3}} \right )|_D.
\end{eqnarray}
Below, we are going to show that 
once this term is added to the VY action (\ref {Action}),
the following fields become dynamical:
\begin{itemize}

\item The $B$ field propagates and it represents
one massive real scalar degree of freedom
(identified later with the scalar glueball).

\item The three-form potential $C_{\mu\nu\alpha}$, which becomes massive,
also propagates. It represents one physical  degree of freedom 
(identified with the pseudoscalar glueball). 

\item The complex field $A$,  being 
decomposed into parity eigenstates,   describes the massive gluino-gluino 
scalar and pseudoscalar mesons.

\item $\chi$ and $\Psi$ describe
the massive gluino-gluon fermionic bound states. 

\end{itemize}
Relations between
masses of these states will be given in the next section.  
\vspace{0.3in} \\
{\bf 3. The mass spectrum } 

Based on the arguments given in the previous section one can write
down the effective Lagrangian  for the lowest-spin multiplets 
of the $N=1$ SUSY YM theory in the following form
\begin{eqnarray}
{\cal L}={1\over \alpha} (S^{+}S)^{1/3}|_D+ 
\gamma [( S \log {S\over \mu^3}-S)|_F+{\rm h.c.}]+   
{1\over \delta} \left ( -{U^2\over (S^+S)^{1/3}} \right )|_D,
\label{NewA}
\end{eqnarray}
where $\alpha~{\rm and}~\delta$ are arbitrary positive constants. 
One obtains the VY Lagrangian in the limit $\delta \rightarrow \infty$.
In general, higher powers of $U$ (and derivatives) 
can also be added to this Lagrangian.
Those terms would introduce new quartic, quintic  and other  higher 
interaction terms. However, the quadratic part of the action which
defines masses will not be affected. 
In that respect, the effective Lagrangian (\ref {NewA}) 
can be treated as the  one describing small 
perturbations of fields about a vacuum state\footnote{Notice that
one can write the Lagrangian  (\ref {NewA})  
in terms of the $U$ field alone. However, for  
clearity of presentation we leave  the VY part of the Lagrangian 
in the original form.}. 

Let us determine the SUSY vacuum state defined by the 
Lagrangian (\ref {NewA}). The potential for the model is 
a complicated function of the variables present in the $U$ 
superfield. After integration over the auxiliary $\Sigma$ field,
the bosonic part of the potential is  
\begin{eqnarray}
V= {2\over \delta (16)^2} {|\phi |^6+\mu^6-2\mu^3|\phi |^3 {\rm cos} 3\rho
\over |\phi |^2}+ {3 \over \delta (48)^2}{ C^2_{\mu\nu\tau}
\over |\phi |^2}+ \nonumber \\
+{9\alpha |\phi |^4\over 4} {1\over 1- {\alpha\over \delta} 
{B^2\over |\phi |^4}} \left ({B\over 24  \delta |\phi |^2}
\{1+ { 2 \mu^3\over |\phi |^3}{\rm cos} 3\rho \}-3\gamma \log {|\phi |^2\over
\mu^2}\right )^2,    
\label{Potential}
\end{eqnarray}
where we introduced the notations  $\phi\equiv A^{1/3}$ and 
$\rho\equiv {\rm arg}\phi$. 

In order to find the vacuum state one
should find the absolute minimum of the potential (\ref {Potential}).
Since we are dealing with a supersymmetric model, the value
of the potential in that minimum has to be zero. 
As a  result of Lorentz invariance the VEV of the three-form field 
is zero, i.e. $\langle C_{\mu\nu\tau} \rangle=0$. 
The VEV of $Q$ is also zero due to the CP invariance of
the model. 
Further calculations are tedious
and we will not present them here. After some algebra one finds that
the only global,  CP invariant  minimum of the potential (\ref {Potential})
is given by:
$\langle \phi \rangle =\mu$, 
$\langle B \rangle =\langle C \rangle = \langle \rho \rangle=0$. 
The effective Lagrangian (\ref {NewA}) describes small 
perturbations of fields about the vacuum state defined by these VEV's.  

We would like to make a comment  here. 
It deals with  the region  of validity of
the potential  (\ref {Potential}) (i.e., of the Lagrangian  (\ref {NewA})).
The expression  (\ref {Potential})
can be a correct potential for a supersymmetric model  if 
$\delta |\phi|^4 > \alpha B^2$. Since the fields  $\phi$ and $B$  
describe perturbations  about the values  $\phi =\mu$ and $B=0$ 
respectively, the inequality  above is 
satisfied for small perturbations of both fields. 
The singularity in the potential at $\delta |\phi|^4 =\alpha B^2$
indicates that for large perturbations the higher dimensional
terms omitted in  (\ref {NewA}) should become important.
As we mentioned above,  we are mainly concerned with the mass 
spectrum of the model which can be studied using small perturbations 
about the ground state, so that the approximation 
given in (\ref {NewA}) is good enough  for our goals
\footnote{The potential (\ref {Potential}) 
describes a SUSY minimum in the field space.
The potential well has  infinitely high walls, at 
values of the fields satisfying  $\delta |\phi|^4 =\alpha B^2$.}.

The mass spectrum for the bosonic part of 
the model can be calculated from the potential (\ref {Potential}).
However, we find it more illuminating to write down the 
quadratic part of the Lagrangian (\ref {NewA}). Then physical 
masses can be computed  as  eigenvalues of the
corresponding quadratic forms. 

Let us rewrite the expression for the Lagrangian (\ref {NewA})
in terms of component  fields keeping only the quadratic terms:
\begin{eqnarray}
{\cal L}= -{1\over 2} \partial_\mu s  \partial_\mu s-
{1\over 2} \partial_\mu p  \partial_\mu p + {1\over 6}C_{\mu\nu\tau}
\partial^2 C^{\mu\nu\tau}-{1\over 2} \partial_\mu B  \partial_\mu B
+i\partial_\mu {\bar \zeta} {\bar
\sigma}^\mu \zeta +i \partial_\mu {\bar \chi} {\bar \sigma}^\mu \chi+
\nonumber \\
\Sigma^2 -{3\over 8} \sqrt {\alpha\over \delta} \mu \Sigma B-
{9 \alpha \over \delta (16)^2}\mu^2 s^2 -{9 \alpha \over \delta
(16)^2}\mu^2 p^2- {27 \alpha \over (48)^2 \delta}\mu^2 C_{\mu\nu\tau}^2
-{9 \alpha \gamma \over 2}\mu {\zeta^2}- {9 \alpha \gamma \over 2}\mu {\bar
\zeta}^2
\nonumber \\
-{i 3 \sqrt{2} \over 16}\sqrt {\alpha\over \delta}\mu \chi \zeta+
{i 3 \sqrt{2} \over 16}\sqrt {\alpha\over \delta}\mu {\bar \chi}
{\bar \zeta}+9\sqrt{2}\alpha \gamma \mu s\Sigma-9\sqrt{2}\alpha\gamma 
 \mu pQ +...
\label{Components}
\end{eqnarray}
where dots stand for cubic and higher dimensional interaction terms.
The fields which appear in eq. (\ref {Components}) are related to
the original fields of eq. (\ref {U}) 
\begin{eqnarray}
A^{1/3}=\phi\equiv \mu+ {\sqrt{\alpha \over 2}}(s+ip),~~~~
\zeta \equiv {\Psi \over 3\sqrt{\alpha} A^{2/3}}. \nonumber 
\end{eqnarray}
Besides that, in eq. (\ref {Components}) 
we performed the following rescaling of the variables 
\begin{eqnarray}
\Sigma \rightarrow 3\sqrt{\alpha}\mu^2 \Sigma,~~~
C\rightarrow 3\sqrt{\alpha}\mu^2 C, ~~~\chi\rightarrow 
\sqrt{\delta} \mu \chi,~~~B \rightarrow \sqrt{\delta} \mu B. 
\end{eqnarray}
The $s$ and $p$ fields, as they stand in the Lagrangian 
(\ref {Components}),   describe respectively the scalar and
pseudoscalar gluino-gluino excitations. The $C$ field, being
a massive three-form potential that propagates one physical 
degree of freedom, 
describes the pseudoscalar glueball. The $\Sigma$ field can be
integrated out by  means of the equation of motion:
\begin{eqnarray} 
\Sigma ={3\over 16} \sqrt {\alpha\over \delta}\mu B-{9 \sqrt{2}\over
2}\alpha \gamma \mu s.
\label{Sigma}
\end{eqnarray}
In accordance with this expression, 
the $B$ field which is left in the effective Lagrangian
describes a mixed state of the scalar glueball (former
$\Sigma$ field ) and the 
scalar gluino-gluino bound state $s$. 

If the VY superpotential were neglected for some reason 
in the expression (\ref {NewA}),
then the two Weyl fermion fields $\chi$ and $\zeta$ would combine
together to form  one Dirac massive bispinor. However, the presence 
of the VY superpotential yields  an additional contribution  to the 
mass term of the $\zeta$ field.  Thus  $\chi$ and $\zeta$ cannot be
treated as the components of one Dirac spinor. Instead one is left with 
two Weyl fermions describing two different  spin 1/2 massive states. 
In general, the form of the bilinear terms in the Lagrangian 
(\ref {Components}) suggests that all the physical states of this
theory should occur as mixed states of the initially  pure bound
states of the Lagrangian (\ref {Components}).

Below, we deduce the masses of these mixed physical states.
Let us write down  the mass and mixing terms of the
Lagrangian (\ref {Components}) separately. 
Substituting the expression for the $\Sigma $ field 
(\ref {Sigma}) into the Lagrangian (\ref {Components}) 
one finds the following pairs of variables being mixed with one
another
\begin{eqnarray} 
B - s ~~ {\rm system}:~~~ {9 \alpha \over \delta (16)^2}\mu^2 s^2+
{81 \over 2} \alpha^2 \gamma^2 \mu^2 s^2+{9 \alpha \over \delta 
(16)^2}\mu^2 B^2-
{27 \sqrt {2}\over 16} \sqrt {\alpha \over \delta}\alpha \gamma \mu^2 Bs;
\nonumber \\
C - p ~~ {\rm system}:~~~~{9 \alpha \over \delta (16)^2 }\mu^2 p^2+
{27 \alpha\over (48)^2 \delta}\mu^2 C_{\mu\nu\tau}^2+
{9\sqrt{2}\over 6}\alpha \gamma  \mu p \varepsilon_{\mu\nu\tau\sigma}
\partial^\mu C^{\nu\tau\sigma};~~~~~~~~~~
\nonumber \\
\chi - \zeta ~~ {\rm system}:~~~~{9 \alpha \gamma \over 2}\mu {\zeta^2}+
{9 \alpha \gamma \over 2}\mu {\bar
\zeta}^2+{i 3 \sqrt{2} \over 16}\sqrt {\alpha\over \delta}\mu \chi \zeta-
{i 3 \sqrt{2} \over 16}\sqrt {\alpha\over \delta}\mu {\bar \chi}
{\bar \zeta}.~~~~~~~~~~~~
\end{eqnarray}

In order to find the  physical masses one must 
diagonalize the corresponding mass matrices. 
Concentrate, for instance, on the first row of these expressions 
which describes the mixed state of scalar meson, $s$, and scalar
meson-glueball, $B$. The former gets mass both from the superpotential and
$U^2$-term while the latter gets mass only from the $U^2$-term.  When the mixing
term is switched on, the initially heavier state ($s$) gets even  heavier,
and the initially lighter state ($B$) becomes even lighter than they were
originally.  Performing the diagonalization, 
one finds that  the physical eigenstates are  mixed states with 
the following mass eigenvalues
\begin{eqnarray}
{1\over 2} m^2_{\rm H}={9 \alpha \over \delta (16)^2 }\mu^2+
{81 \over 4} \alpha^2 \gamma^2 \mu^2 +{81 \over 4} \alpha^2 \gamma^2 \mu^2
\sqrt{ 1+ {1 \over 288}{ \alpha\over  \delta} {1\over
(\alpha\gamma)^2} },
\label{up}
\end{eqnarray}
and 
\begin{eqnarray}
{1\over 2} m^2_{\rm L}={9 \alpha \over \delta (16)^2 }\mu^2+
{81 \over 4} \alpha^2 \gamma^2 \mu^2 -{81 \over 4} \alpha^2 \gamma^2 \mu^2
\sqrt{ 1+ {1 \over 288}{ \alpha\over  \delta} {1\over
(\alpha\gamma)^2} }.
\label{down}
\end{eqnarray}
Here, the subscript ``H'' refers to the heavier state ${\tilde s}$
which, without
mixing,  would have been a 
 pure gluino-gluino bound state (the $s$ particle). 
``L'' refers to the lighter state ${\tilde B}$  ($B$ in the absence 
of mixing). 

Explicit calculation shows that the $C-p$ and $\chi-\zeta$ 
systems possess exactly the same properties. Namely, the heavier 
mass eigenstates   ${\tilde p},~~  {\tilde \zeta}$  
acquire  the mass squared eigenvalues  given by the expression 
for $m^2_{\rm H}$, and the lighter eigenstates  ${\tilde C},~~  {\tilde
\chi}$ have  mass squared eigenvalues  equal to 
$m^2_{\rm L}$.
Thus, as excpected, the physical states form two multiplets, one with mass
$m^2_{\rm H}$ and the other with mass $m^2_{\rm L}$.

Let us discuss various limits of eqs. (\ref{up}) and (\ref{down}).
Suppose $\alpha\rightarrow 0$. In that limit the superpotential
and the $U^2$  terms can be neglected in the Lagrangian
(\ref {NewA}). The K\"ahler potential, which would be the only term left in 
the expression  (\ref {NewA}), would yield  only kinetic terms for the excitations.
Thus all those states would be massless. This is in agreement with the expressions
(\ref{up}) and (\ref{down}) which turn into zero as $\alpha\rightarrow
0$. 

In the $\gamma\rightarrow 0$ limit the superpotential disappears. Thus, the
mass terms  for  the physical states come only from the $U^2$ term in  the
Lagrangian  (\ref {NewA}). As a result, all the masses are expected to be
equal and there is  no mixing  between pure gluonic
and fermionic states described above. Also, as noted earlier, the $\chi$
and $\zeta$ Weyl fermions come together to form 
one massive Dirac bispinor. The limit $\gamma\rightarrow 0$ is
physically equivalent to the limit $\delta \rightarrow 0$; in both
cases the superpotential can be neglected in comparison with 
the $U^2$ term. 

Finally, let us consider the $\delta \rightarrow \infty$
limit. One can neglect the $U^2$ term in that case. As a result
one rederives  the VY Lagrangian  with the spectrum given by the 
second term in eq. (\ref{up}) (or eq.  (\ref{down})) multiplied
by the factor of two. No glueballs are 
incorporated in the effective theory in that limit.  
\vspace{0.3in} \\
{\bf Summary and Discussion }

We have shown here how to generalize the Veneziano-Yankielowicz effective
action for $N=1$ supersymmetric Yang Mills theory to include composite
operators corresponding to pure gluonic, and not exclusively
gluino-containing, bound states.  We accomplish this by embedding the chiral
multiplet of anomalies into a larger three-form tensor supermultiplet and
adding an extra term to the VY effective action.  The extra term is necessary
in order  to retain the variables corresponding to glueballs  as  dynamical
fields in the effective Lagrangian. 

Studying the potential of the model, we find  that
the physical eigenstates fall into the  two different 
``multiplets" with masses given by eqs. (\ref {up}) and (\ref {down}).  
Neither of them contain pure gluino-gluino, gluino-gluon or
gluon-gluon bound states.  Instead, the physical excitations are 
mixed states of these composites.  The heavier set of states contains: 
\begin{itemize}
\item A pseudoscalar meson, which without mixing reduces to the $0^{-+}$
gluino-gluino bound state (the analog of the QCD $\eta'$ meson).
\item A scalar meson that without mixing is an $l=1$ $0^{++}$ gluino-gluino 
excitation.
\item A mixed fermionic gluino-gluon bound state. 
\end{itemize}
These heavier states become the chiral supermultiplet described by the VY
action in the limit that the additional term we have added to the effective
Lagrangian is  removed.  The new states which appear as a result of our 
generalization forms a lighter multiplet:
\begin{itemize}
\item A scalar meson, which for small mixing becomes a $0^{++}$ ($l=0$)
glueball. 
\item A pseudoscalar state, which for small mixing is identified as a
$0^{-+}$ ($l=1$) glueball.
\item A mixed fermionic gluino-gluon bound state. 
\end{itemize}

We call the reader's attention to an interesting feature of the
effective action introduced here.  Although the physical states fall into
multiplets whose $J^P$ quantum numbers correspond to two chiral 
supermultiplets,
the action is not written in terms of two chiral supermultiplets.
Another  representation of SUSY is used, as explained in
section 2.  In particular, the pseudoscalar glueball in this approach is
described by the only physical component of the massive three-form potential
$C_{\mu\nu\alpha}$. The field strength of that potential couples to the
pseudoscalar gluino-gluino bound state as it would couple to the $\eta'$
meson in QCD. 

Masses of the physical states depend on two independent mass parameters
($\frac{\alpha}{\delta} \mu$ and ${\alpha}{\gamma} \mu$) formed from constants
occuring in the effective Lagrangian.  However the result that
the lighter multiplet contains the predominantly glueball excitation is
independent of the values of these parameters.  This prediction
receives some support from recent lattice studies of SUSY YM theory.
Although the lattice results for the mass spectrum are available 
only away from the SUSY point, the general tendency is that a state which
is mostly a glueball is lighter than the state which is mostly a gluino-gluino
composite \cite {Lattice}\footnote{We refrain from using quenched
approximation results \cite{lattice:cernrome} to obtain information on the
supersymmetric theory because in SUSY Yang-Mills theory all composite masses
are proportional to the gluino condensate, as can be seen explicitly from
our results.  In the absence of dynamical gluinos, masses can only be
proportional to the SUSY breaking condensate $\langle G_{\mu \nu} G^{\mu
\nu}\rangle $.}.  
More detailed lattice results would in principle permit fixing the
combinations $\frac{\alpha}{\delta} $ and ${\alpha}{\gamma}$ 
which are undetermined within the effective Lagrangian approach.
However determining the masses of the two 
multiplets is likely to be difficult:
if mixing is small they are nearly degenerate and if mixing is large
they are in general both excited by a given source.
\vspace{0.3in} \\
\noindent{\bf Acknowledgments }

The authors are grateful to J. Gates for emphasizing the possible importance
of non-standard representations of SUSY, one of which proved essential in
our construction. The work of M.S. was supported by the NSF grant
PHY-94-23002.


\begin{thebibliography}{99}

\bibitem{S} N. Seiberg, ``Exact Results On The Space Of 
Vacua Of Four-Dimensional SUSY Gauge Theories'',  
Phys. Rev. D49(1994)6857.


\bibitem{SW} N. Seiberg, E. Witten, ``Electric-Magnetic Duality,
Monopole Condensation, And Confinement In $N=2$  Supersymmetric
Yang-Mills Theory'',  
Nucl.Phys. B426(1994)19, Erratum-ibid. B430(1994)485;  
``Monopoles, Duality And Chiral Symmetry Breaking In  $N=2$   
Supersymmetric QCD'', Nucl.Phys. B431(1994)484. 


\bibitem{VY} G. Veneziano, S. Yankielowicz, ``An Effective Lagrangian
For The Pure $N=1$ Supersymmetric Yang-Mills Theory'', Phys. Lett. 
113B(1982)231. 


\bibitem{WessZumino} S. Ferrara, B. Zumino, 
``Transformation Properties Of The Supercurrent``,
Nucl. Phys. B87 (1975)207.

\bibitem{WessBagger} J. Wess, J. Bagger, ``Supersymmetry And
Supergravity'', 2nd ed., Princeton University Press, Princeton,  1992.


\bibitem{beta} V.A. Novikov, M.A. Shifman, A.I. Vainshtein, 
V.I. Zakharov, ``Exact Gell-Mann - Low Function Of Supersymmetric 
Yang-Mills Theories From Instanton Calculus'', Nucl. Phys. B229(1983)381.


\bibitem{Shore} G.M. Shore, ``Constructing Effective Actions 
For $N=1$ Supersymmetry Theories: Symmetry Principles And
Ward Identities'', Nucl. Phys. B222(1983)446.


\bibitem{Seiberg} K. Intriligator, R.G.  Leigh, N. Seiberg, 
``Exact  Superpotentials In  Four-Dimensions``,
Phys. Rev. D50(1994)1092. \\
K. Intriligator, N. Seiberg, 
``Lectures  On  Supersymmetric  Gauge  Theories  And  Electric -
Magnetic  Duality '', Nucl.Phys.Proc.Suppl.45BC(1996)1; 
hep-th/9509066.  

\bibitem{KShifman} A. Kovner, M. Shifman, ``Chirally Symmetric Phase
Of Supersymmetric Gluodynamics'', Phys. Rev. D56(1997)2396. 

\bibitem{Condensate} V.A. Novikov, M.A. Shifman, A.I. Vainshtein, 
V.I. Zakharov, ``Supersymmetric Instanton Calculus (Gauge Theories
With Matter)'', Nucl. Phys. B260(1985)157;\\
M.A. Shifman, A.I. Vainshtein, ``On Gluino Condensation
In Supersymmetric Gauge Theories. $SU(N)$ And $O(N)$ Groups'', 
Nucl. Phys. B296(1988)445. 

\bibitem{mm} M. Schwetz and M. Zabzine, "Gaugino Condensate And
Veneziano-Yankielowicz Effective Lagrangian", hep-th/9710125.

\bibitem{cm} C. Csaki and H. Murayama, "Discrete Anomaly Matching",
hep-th/9710105.

\bibitem{SVZ} M.A. Shifman, A.I. Vainshtein, ``Solution Of The Anomaly
Puzzle In SUSY Gauge Theories And The Wilson Operator Expansion'', 
Nucl. Phys.  B277(1986)456.


\bibitem{GoldstoneSalamWeinberg} 
J. Goldstone, A. Salam, S. Weinberg, Phys. Rev.
127(1962)965.

\bibitem{Schechter} J. Schechter, ``Effective Lagrangian With Two
Color Singlet Gluon Fields'', Phys. Rev. D21(1980)3393.

\bibitem{MigdalShifman}  A.A. Migdal, M.A. Shifman, ``Dilaton Effective
Lagrangian In Gluodynamics'', Phys. Lett. 114B(1982)445.

\bibitem{Gabad} G. Gabadadze, ``Modeling The Glueball Spectrum By
A Closed Bosonic Membrane'', hep-ph/9710402.

\bibitem{OgievetskyMezincescu} V.I. Ogievetskii, L. Mezincescu, 
``Boson-fermion Symmetries And Superfields'', 
Sov. Phys. Usp. 18(1975)960.

\bibitem{Landafshitz} V.B. Berestetskii, E.M. Lifshitz,
L.P. Pitaevskii, ``Quantum Electrodynamics'', Nauka, Moscow, 1989.
 
\bibitem{NSM}  N. Evans, S.D.H. Hsu, M. Schwetz, ``Lattice Tests Of 
Supersymmetric Yang-Mills Theory ?'',  hep-th/9707260. 

\bibitem{V} P. Di Vecchia, G. Veneziano, ``Chiral Dynamics In The
Large $N_c$ Limit'', Nucl. Phys. B171(1980)253.

\bibitem{Gates} S.J. Gates Jr., ``Super  P Form 
 Gauge Superfields'', 
Nucl. Phys. B184(1981)381.

\bibitem{OvrutWaldram} P. Binetruy, F. Pillon, G. Girardi, R. Grimm,
``The  Three  Form  Multiplet  In  Supergravity''
Nucl. Phys. B477(1996)175;
\\B.A. Ovrut, D. Waldram, ``Membranes And
Three-form Supergravity'', hep-th/9704045.

\bibitem{GatesGRZ} S.J. Gates Jr., M.T. Grisaru, M. Rocek, W. Siegel,
``Superspace or One Thousand And One Lessons In Supersymmetry'', 
Benjamin/Cummings, Massachusetts, 1983.
\bibitem{Derendinger} C.P. Burgess, J.-P. Derendinger,
F. Quevedo, M. Quiros, ``Gaugino Condensates and Chiral-Linear
Duality: an Effective Lagrangian Analysis'', Phys. Lett. B348(1995)428.

\bibitem{Lattice} I. Montvay, ``SUSY On The Lattice'',
hep-lat/9709080; Talk given at Lattice 97: 15th
International Symposium on Lattice Field Theory, 
Edinburgh, Scotland, 22-26 Jul 1997;
G. Koutsoumbas, I. Montvay, A. Pap, K. Spanderen, D. Talkenberger, J.
Westphalen, ``Numerical Study Of SU(2) Yang-Mills 
Theory With Gluinos'', hep-lat/9709091.

\bibitem{lattice:cernrome} A. Donini, M. Guagnelli, P. Hernandez,
A. Vladikas, ``Towards $N=1$ Super-Yang-Mills on the Lattice'',
hep-lat/9710065.


\end{thebibliography}
\end{document}